\documentclass[twocolumn,british,useAMS,usenatbib]{mn2e}
\usepackage{mathptmx}

\usepackage[T1]{fontenc}
\usepackage[utf8]{inputenc}
\usepackage{babel}
\usepackage{color}
\usepackage{textcomp}
\usepackage{amsmath}
\usepackage{amssymb}
\usepackage{graphicx}
\usepackage[authoryear]{natbib}
\usepackage{subscript}

\makeatletter

\usepackage{colortbl}
\definecolor{LightGray}{gray}{0.9}

\title[Near-parabolic comets observed in 2006--2010. II] {Near-parabolic comets observed in 2006--2010. II. Their past and future motion under the influence of the Galaxy field and known
nearby stars. }
\author[P.A. Dybczy\'{n}ski \& M. Kr\'olikowska]{Piotr A. Dybczy\'{n}ski$^{1}$\thanks{E-mail: dybol@amu.edu.pl} \& Ma\l{}gorzata Kr\'olikowska$^2$\thanks{E-mail:mkr@cbk.waw.pl}\\
$^1$Astronomical Observatory Institute,
A.Mickiewicz Univ., S\l{}oneczna 36, 60-286 Pozna\'{n}, Poland;\\
$^2$Space Research Centre of the Polish Academy of Sciences,
Bartycka 18A, 00-716 Warsaw, Poland.}

\date{Accepted 2015 January 3.  Received 2014 December 10; in original form 2014 September 1}
\pagerange{\pageref{firstpage}--\pageref{lastpage}} \pubyear{2015}

\makeatother

\begin{document}

\maketitle

\label{firstpage} 
\begin{abstract}
In the first part of this research we extensively investigated
and carefully determined osculating, original (when entering Solar
system) and future (when leaving it) orbits of 22~near-parabolic
comets with small perihelion distance ($q_{\rm osc}<3.1$\,au),
discovered in years 2006-2010. Here, we continue this research with
a detailed study of their past and future motion during previous and
next orbital periods under the perturbing action of our Galactic environment.
At all stages of our dynamical study, we precisely propagate in time
the observational uncertainties of cometary orbits. For the first
time in our calculations, we fully take into account individual perturbations
from all known stars or stellar systems that closely (less than 3.5\,pc)
approach the Sun during the cometary motion in the investigated time
interval of several million years. This
is done by means of a direct numerical integration of the N-body system
comprising of a comet, the Sun and 90 potential stellar perturbers.
We show a full review of various examples of individual stellar action
on cometary motion. We conclude that
perturbations from all known stars or stellar systems do not change
the overall picture of the past orbit evolution of long-period comets
(LPCs).The future motion of them might be seriously perturbed during the 
predicted close approach of Gliese\,710 star but we do not observe significant
energy changes. The importance of stellar perturbations is tested
on the whole sample of 108~comets investigated by us so far and
our previous results, obtained with only Galactic perturbations included,
are fully confirmed. We present how our results can
be used to discriminate between dynamically new and old near-parabolic comets
and discuss the relevance of the so-called Jupiter-Saturn barrier phenomenon. Finally, we show
how the Oort spike in the $1/a$-distribution of near-parabolic comets
is built from both dynamically new and old comets. We
also point out that C/2007~W1 seems to be the first serious candidate for interstellar provenience.
\end{abstract}


\section{Introduction}

This is a second part of our extensive investigation of dynamics of
long-period comets discovered during the period of 2006--2010. In
Part~I \citep{kroli-dyb:2013}, we described in detail the astrometric
data and their tailored numerical treatment which allowed us to obtain
the most accurate osculating orbits, with nongravitational forces
taken into account wherever possible. Then we propagated these orbits
forward and backward up do the distance of 250 au from the Sun to
obtain\emph{ original} (i.e. before entering the planetary perturbations
zone) and \emph{future} (i.e. when leaving it) orbits. All planetary
perturbations were strictly taken into account as well as relativistic
effects. We also took into account the existence of nongravitational
accelerations in the motion of studied comets, and in up to half of
the cases we were able to obtain satisfactory nongravitational orbits.
In some cases of well-observed LPCs, we used subsets of astrometric
observations, omitting those near perihelion, where violent cometary
activity can disturb the motion noticeably.

At the stage of the osculating orbit, we augmented the nominal orbit
of a comet with 5000 of its clones (virtual comets, VCs) which all
represent astrometric observations on the same level as the nominal
orbit does. This allowed us to control the observational uncertainties
at each step of our dynamical studies. The results and a deep discussion
of all the above may be found in Part~I.

Here, we present the result of the natural next step of cometary orbits
investigation, namely examination of their evolution for one orbital
period to the past and future. Some elements of our numerical methods
are presented in Section~\ref{sec:Methods-of-calculation}. Similar analysis 
for two different samples of near-parabolic comets we published
earlier, see \citet{kroli-dyb:2010}, hereafter Paper~I, \citet{dyb-kroli:2011},
hereafter Paper~II, and \citet{kroli-dyb:2012} (Paper~III). In
these papers, we generally ignored stellar perturbations, according
to our early tests \citep{dyb-hist:2001} and more detailed examinations
taken later \citep{dyb-hab3:2006}.

However, in some recent papers the authors pointed out, that stellar
perturbations play a significant role in long term simulations of
cometary dynamics (see for example: \citealp{fouchard-r-f-v:2011})
and almost simultaneously a new and comprehensive compilation of modern
stellar data have been published -- XHIP catalogue \citep{anderson_francis:2011}.
This catalogue, prepared mainly on the basis of the new reduction
of Hipparcos measurements \citep{vanleeuwen:2007,vanleeuwen:2011}
and augmented with over 46 thousand radial velocities from various
modern sources, can serve as ideal tool for searching for potential
stellar perturbers of the observed LPCs.

For that reason, we decided to include stellar perturbations from
known stars in our dynamical model of near-parabolic comets motion.
We performed an extensive search for potential stellar perturbers
and found about 90~stars which visited solar neighbourhood in the
recent past or will visit it in the near future. The details are given
in Section~\ref{sec:Stellar-perturbations}.  Please note that we 
use the most up to date stellar data compilation. While we are rather 
convinced that none of the slow moving and massive stars visiting 
recently the solar neighbourhood remain unknown we must admit that 
any new findings might happen to  change our results and conclusions 
for a particular comet.  The ongoing mission GAIA will greatly improve 
the accuracy of the available data on nearby stars and may discover 
a new potential stellar perturber of cometary motion in next years.

In Sections~\ref{sec:Past-motion} and \ref{sec:Future-motion} we
present an analysis of past and future dynamical evolution of a complete
sample of small-perihelion LPCs discovered during the period of 2006-2010.
We show how these results can be used to discriminate between dynamically
old and new LPCs. We call a comet dynamically new if the observed
orbit was not perturbed by planets during its previous perihelion
passage. In other words this comet makes it first visit in the solar
neighbourhood. The rest of LPCs we call dynamically old or uncertain.
For the uncertainty range between the dynamically new and old comets,
we adopted here the range of previous perihelion distance from 10\,au
to 20\,au. This is a more cautious condition than in Paper~II, where
we generally assumed that a comet is dynamically new when its previous
perihelion was greater than 15\,au from the Sun.

Special attention was given to the comparison of the results obtained
with and without stellar perturbations to verify their importance.
This is discussed in Section \ref{sec:Examples-of-the}.

\section{Methods of calculations in brief}

\label{sec:Methods-of-calculation}

\vspace{0.2cm}

The starting point for the calculations described in this paper is
an original/future orbit of each comet, when past/future evolution
is examined, respectively. In fact, we use not only the nominal barycentric
orbit at 250\,au from the Sun but also 5\,000 clones of it, together
constituting a swarm of 5\,001 original/future barycentric orbits
for each examined comet. The procedure of determining the osculating
orbits from astrometric observations, the cloning algorithm and the
method of obtaining original and future swarms of orbits are described
in detail in Part~I.

\emph{Original} and \emph{future} LPCs orbits were followed numerically
taking into account Galactic perturbations (where both disk and central
terms were included) and any possible stellar perturbation from known
stars, their selection and treatment are described in detail in the
next section. Generally, we stop our calculation at a previous or
next perihelion but for hyperbolic or extremely elongated elliptic
orbits in the swarm of VCs we apply so called ``escape limit'' of
120\,000\,au. The final orbits obtained from these calculations
we call \emph{previous} and \emph{next}.

In this paper we call a comet (or more precisely each individual VC)
as returning {[}R{]} if it goes no further than 120\,000~au from
the Sun. All other comets (or VCs) are named escaping {[}E{]}, but
among them we separately count escapes along hyperbolic {[}H{]} orbits.

For the detailed description of the dynamical model as well as its
numerical treatment the reader is kindly directed to Paper~I. Basing
on the conclusions from that work we used both, Galactic disk and
Galactic centre terms in all calculations. All the parameters of the
Galactic gravity field are kept unchanged, including the local disk
mass density, $\rho=0.100$ M$_{\odot}/pc^{3}$.  The recently quoted 
value \citep{irrgang_et_al:2013} equals 0.102 ± 0.010. As we showed in 
one of our previous papers \citep{kroli-dyb:2010} even assuming the 50 
per cent uncertainty it influences our results on a less than 1 per cent 
level.   The only one but important modification of the dynamical 
model is the incorporation of any possible  known  stellar 
perturbations (Section~\ref{sec:Stellar-perturbations}).

Dynamical behaviour of VCs within an individual swarm forced the rules
of stopping the numerical integration. We decided to treat each VC
in a given swarm as uniformly as possible, thus, we distinguish between
the following three cases. \vspace{-0.2cm}

\begin{enumerate}
\item When all VCs of a particular comet were returning then all of them
were stopped in two ways: \vspace{-0.2cm}

\begin{itemize}
\item at previous/next perihelion (basic variant), 
\item simultaneously when the nominal VC reached previous/next perihelion
(synchronous variant). 
\end{itemize}

\vspace{-0.2cm}
 The reason of introducing the synchronous variant is that for some
purposes it more convenient to have all VCs stopped at the same moment
of time, which is not satisfied in the basic variant. \vspace{0.2cm}

\item When all VCs were escaping then the calculation was stopped synchronously,
at the moment when the fastest VC crossed the escape limit (usually
very close to 120\,000~au). \vspace{0.2cm}

\item When a swarm of VCs consists of both returning and escaping VCs then
we also decided to stopped them in two ways: \vspace{-0.4cm}

\begin{itemize}
\item the returning part was stopped at previous/next perihelion and the
rest (escaping ones) when the fastest escaping VC crossed the escape
limit (mixed variant). 
\item all VCs (both returning and escaping) were stopped at the moment when
the fastest VC crossed the escape limit (synchronous variant). 
\end{itemize}

\vspace{-0.2cm}
 Similarly, as in the first case, this synchronous variant were used
to examine more homogeneous orbital elements distributions, i.e. when
all VCs were stopped at the same epoch.

\end{enumerate}

\section{Stellar perturbations}

\label{sec:Stellar-perturbations}

To obtain  a more reliable picture of the past and future
motion of the investigated comets, we decided to include stellar perturbations
in our dynamical model of the distant cometary motion. For this purpose,
we completely revised our list of potential stellar perturbers acting
on the LPCs motion in large heliocentric distances. While the detailed
analysis of all available stellar data and their current accuracy
will be presented in a future paper (Dybczyński and Berski, in preparation),
now we use a recently published compilation called XHIP \citep{anderson_francis:2011}
as a source of 6-D stellar data. At the moment, we completely rely
on the results presented in this catalogue and accept all data refinement
rules adopted by the authors. 
 As it concerns the completeness of the available stellar data the objections expressed by 
\citet{rickman_ffv:2012} are based on a simulated stellar passage frequency obtained from the 
estimations done by \citet{garcia-sanchez:2001}. Their estimations should be treated as the 
'long time mean'. It is easy to guess that this frequency will vary significantly with time. 
As a simple proof  lets consider stars in the sphere of 20 pc radius around the Sun. 
A relatively fast star, with the total velocity of 40 km/s moves about 20 pc in 0.5 million years. 
None of the known stars pass as close as 1 pc  in 1 million years centred at present epoch. 
Note, that our knowledge of stars in 20 pc range is rather complete except  spectral type M 
and smaller \citep{jehreiss_wielen:1997} . The Garcia-Sanchez estimation predicts 2-3 stars in each 
million years but currently we cannot find any. As we go to slower and more massive stars, 
which are most important perturbers of cometary motion our knowledge extends to larger time 
intervals, comparable with one orbital period of investigated long-period comets.

However, the reader should be warned that any new discovery of a new, strong 
stellar perturber, while of little probability in our opinion, may change our results and 
conclusions concerning one or another particular comet.

\subsection{The model}

\label{sec:Stellar-perturbations-model}

We extracted from the XHIP catalogue all stars with the calculated
proximity distance (Dmin in XHIP) less than 3.5\,pc from the Sun.
There are 96 such stars in this catalogue. One of them (HIP 72511)
must be excluded due to the evident parallax error. Taking into account
its spectral type (M1) and the apparent magnitude (m$_{\textrm{v}}$=11.7)
its parallax must be much smaller, perhaps, the value of 30\,mas
presented in some older sources (see for example \citealp{auer:1978})
or 40 mas in Gliese and Yale catalogues (\citealp{gliese_jahreiss:1993,yale-cat:1995})
is much closer to the reality than the Hipparcos value of 248\,mas.
Moreover, if this large parallax had been real, this star would have
been included in the RECONS list of the 100 nearest stellar systems
\footnote{http://www.chara.gsu.edu/RECONS/TOP100.posted.htm%
}, which is not the case.

\noindent Additionally, we decided to replace individual members of
four stellar systems with their centres of mass. We merged HIP 13769
with HIP 13772 (GJ 120.1 AB+C), HIP 26369 with HIP 26373 , HIP 70890
with HIP 71681 and HIP 71683 (Proxima + $\alpha$~Centauri~A+B)
and finally HIP 104214 with HIP 104217 (GJ 280 A+B).

All of these changes, reduced the sample of the stellar perturbers
to 90 stars or stellar systems. This list we divided into three parts: 
\begin{itemize}
\item 11 current neighbours (group C), i.e. stars that are currently closer
than 3.5 pc 
\item 39 past visitors (group P) - stars that have passed closer than 3.5
pc from the Sun in the past (the oldest such an event among stars
for which we have complete spatial data occurred more than 7 million
years ago) 
\item 40 future visitors (group F) - stars that will make close passage
in the future (the most distant in time event, included for these
stars (again, with complete spatial data available) will occur almost
5 million years from now). 
\end{itemize}
As the method of cometary motion calculation, we chose the numerical
integration of the equation of motion expressed in the rectangular,
heliocentric, Galactic coordinates. We used the popular and well-tested
RA15 routine \citep{everhart-ra15:1985}. We decided to use the so
called ``local Galactic potential'' (see for example \citealp{jimenez_et_al:2011})
acting on both a comet and each stellar perturber. We also include
all N-body mutual interactions between stars (including our Sun) and
their gravitational influence on a comet. As it concerns Galactic
perturbations we used exactly the same model for disk and centre tides
as that described in detail in Paper~I.

\subsection{Overall characteristics of stellar influence on the motion of actual
near-parabolic comets}

\label{sec:Stellar-perturbations-characteristics}

For a preliminary review of stellar action on all LPCs studied by
us so far, we performed a special numerical integration of the past
and future motion of nominal orbits of 22~comets investigated in
Part~I of this study \citealp{kroli-dyb:2013} and, additionally,
of all 86~comets investigated in our previous papers (Paper~I and
Paper~II).

At first, we compared previous and next nominal perihelion distances
for all these 108~comets (except of the next perihelion of C/2010
X1~Elenin because this comet disintegrated near perihelion) with
the values obtained without stellar perturbations.

It appears that in the backward motion no important (especially from
the point of view of discerning dynamically new/old comets) changes
can be observed. The largest observed difference is on the level of
38 per cent: the nominal previous perihelion of C/2006~OF$_{2}$
decreased from 36.2\,au (model without stellar perturbations) down
to 22.5\,au (with the stellar action switched on). The previous perihelion
distances of only two from among 108~near-parabolic comets is slightly
``moved'' across the adopted limits of 10 or 15\,au (used in Paper~II),
i.e. their dynamical status has to be reclassified. Namely, the previous
nominal perihelion distance changed from 9.58\,au to 10.16\,au for
C/1996~E1, and from 15.05\,au to 14.80\,au for C/2001~K3. As a
result both comets should be classified as dynamically uncertain.
None of all 108 comets is moved across the 20\,au threshold value,
this is an additional reason why we adopt this value as defining dynamically 
new comets in this paper.

\begin{table*}
\caption{\label{tab:past_motion_old} The past distributions of swarms of VCs
in terms of returning {[}R{]}, escaping {[}E{]}, including hyperbolic
{[}H{]} VC numbers for dynamically old comets. Previous aphelion (col.~7)
and perihelion (col.~8) distances are described either by a mean
value (in cases where the normal distribution approximation is applicable)
or three deciles at 10, 50 (i.e. median), and 90 per cent. We present
two rows for each comet: the upper row (highlighted by grey shading)
contains the results with the stellar perturbations included and the
second row (presented here for comparison) shows results where stellar
perturbations were ignored in our dynamical calculations. For data
completeness, we include the osculating perihelion distance in the
third column; the last column presents the value of $1/{\rm a}_{{\rm prev}}$.}

\begin{tabular}{llccccccr}
\hline 
 & model  & $q_{{\rm osc}}$  & \multicolumn{3}{c}{Number of VCs} & $Q_{{\rm prev}}$  & $q_{{\rm prev}}$  & $1/a_{{\rm prev}}$\tabularnewline
Comet  & \& quality  & au  & {[}R{]}  & {[}E{]}  & {[}H{]}  & $10^{3}$au  & au  & $10^{-6}$au$^{-1}$\tabularnewline
 & class  &  &  &  &  &  &  & \tabularnewline
\hline 
{[}1{]}  & {[}2{]}  & {[}3{]}  & {[}4{]}  & {[}5{]}  & {[}6{]}  & {[}7{]}  & {[}8{]}  & {[}9{]}\tabularnewline
\hline 
\rowcolor{LightGray}C/2006 HW\textsubscript{51} Siding Spring  & GR, 1a  & 2.27  & 5001  & 0  & 0  & 38.8 - 42.3 - 46.6  & 2.44 - 2.81 - 3.72  & 47.21 \textpm{} 3.37\tabularnewline
 &  &  & 5001  & 0  & 0  & 38.8 - 42.3 - 46.7  & 2.37 - 2.73 - 3.63  & 47.20 \textpm{} 3.37\tabularnewline
\rowcolor{LightGray}C/2006 K3 McNaught  & NG, 1a  & 2.50  & 5001  & 0  & 0  & 29.8 - 32.7 - 36.2  & 2.34 - 2.35 - 2.39  & 61.11 \textpm{} 4.63\tabularnewline
 &  &  & 5001  & 0  & 0  & 29.8 - 32.7 - 36.2  & 2.35 - 2.36 - 2.45  & 61.11 \textpm{} 4.63\tabularnewline
\rowcolor{LightGray}C/2006 P1 McNaught  & NG, 1b  & 0.17  & 5001  & 0  & 0  & 32.1 - 34.9 - 38.5  & 0.079-0.093-0.106  & 57.20 \textpm{} 4.03\tabularnewline
 &  &  & 5001  & 0  & 0  & 32.1 - 34.9 - 38.5  & 0.073-0.094-0.110  & 57.21 \textpm{} 4.03\tabularnewline
\rowcolor{LightGray}C/2006 Q1 McNaught  & NG, 1a+  & 2.76  & 5001  & 0  & 0  & 39.15 \textpm{} 0.37  & 4.944 \textpm{} 0.090  & 51.09 \textpm{} 0.48\tabularnewline
 &  &  & 5001  & 0  & 0  & 39.15 \textpm{} 0.37  & 4.834 \textpm{} 0.086  & 51.08 \textpm{} 0.48\tabularnewline
\rowcolor{LightGray}C/2007 Q3 Siding Spring  & GR-PRE, 1a+  & 2.25  & 5001  & 0  & 0  & 47.73 \textpm{} 0.60  & 9.59 \textpm{} 0.44  & 41.90 \textpm{} 0.53\tabularnewline
 &  &  & 5001  & 0  & 0  & 47.71 \textpm{} 0.60  & 9.38 \textpm{} 0.43  & 41.92 \textpm{} 0.53\tabularnewline
\rowcolor{LightGray}C/2008 A1 McNaught  & NG-PRE, 1b  & 1.07  & 5001  & 0  & 0  & 16.56 \textpm{} 0.28  & 1.0599 \textpm{} 0.0006  & 120.81 \textpm{} 2.03\tabularnewline
 &  &  & 5001  & 0  & 0  & 16.56 \textpm{} 0.28  & 1.0594 \textpm{} 0.0008  & 120.82 \textpm{} 2.03\tabularnewline
\rowcolor{LightGray}C/2009 K5 McNaught  & GR-PRE, 1a  & 1.42  & 5001  & 0  & 0  & 43.95 \textpm{} 0.53  & 3.70 \textpm{} 0.15  & 45.50 \textpm{} 0.55\tabularnewline
 &  &  & 5001  & 0  & 0  & 43.96 \textpm{} 0.53  & 3.55 \textpm{} 0.15  & 45.50 \textpm{} 0.55\tabularnewline
\rowcolor{LightGray}C/2009 O4 Hill  & GR, 1b  & 2.56  & 5001  & 0  & 0  & 32.1 - 35.7 - 40.3  & 1.69 - 1.97 - 2.15  & 55.94 \textpm{} 4.91\tabularnewline
 &  &  & 5001  & 0  & 0  & 32.2 - 35.7 - 40.3  & 1.58 - 1.86 - 2.05  & 55.96 \textpm{} 4.91\tabularnewline
\hline 
\end{tabular}
\end{table*}

We have also checked all minimal distances between comets (on their
nominal orbits) and stars, finding that the proximities closer than
1\,pc are very rare in the past. In fact only two stars, HIP~30344
and HIP~84263, passed near some of comets investigated by us so far
closer than 1\,pc. The closest passage of HIP~84263 was at a distance
of 0.52\,pc from C/1987~W3, and the closest flyby of HIP~30344
was at 0.85\,pc from C/2006~OF$_{2}$. HIP~30344 is a moderate
perturber ( with mass of 0.79\,M$_{\odot}$ and a velocity of 18.3\,km\,s$^{-1}$)
while HIP~84263 can potentially act much stronger (1.3\,M$_{\odot}$,
$V$=12.6 km\,s$^{-1}$). But even such a
slow and massive star acts rather weakly from a distance greater than
150\,000\,au -- it increases the nominal previous perihelion distance
of C/1987~W3 from 78.9\,au up to just 83.8\,au.

The situation is a bit more complicated when we look at the future
cometary motion under the stellar perturbations. It is a well-known
fact that the star Gliese~710 (HIP~89825) will pass near the Sun
in 1.4~million years at a very close distance. The predicted proximity
distance strongly depends on the 6-D initial data and the method of
a stellar trajectory calculation. Using our model and data
taken from XHIP catalogue, we obtained its closest heliocentric distance
as small as 0.23\,pc, almost equal to the 0.24\,au value from the
linear approximation presented in XHIP. This makes an arbitrarily
close approach of this star to some LPCs with favourable orbit orientations
quite possible to occur. While the mass of GJ~710 is not so big (estimated
to be 0.6\,M$_{\odot}$) its arbitrarily close approach can potentially
perturb cometary orbit drastically. Gliese~710 is the only star from
the tested sample of 90 potential stellar perturbers which can pass
in the future closer than 1\,pc from any of investigated comets.
We registered only few additional evens with the $\alpha$~Centauri
system but almost exactly at the 1\,pc distance.

Therefore, Gliese~710 acts on almost all investigated comets during
their next orbital revolution, except from these which change their
orbits (due to the planetary perturbations during the observed apparition)
into much more bound ones. With the orbital period smaller than 1.3~million
years (semimajor axis less than 12\,000\,au) they cannot experience
any significant perturbation caused by Gliese~710. Comets with larger
semimajor axes received smaller or bigger ``kicks'' from Gliese~710,
which change their future orbits substantially in some cases. However,
from among the 22\,comets studied in this paper only
three LPCs escaping in the future (C/2006~K3, C/2006~L2 and C/2009~O4)
were affected, as well as the widely spread swarm of VCs of C/2007~Q1
(poor quality orbit) of mixed nature (some of the VCs are returning
and some are escaping). The detailed influence of stellar perturbations
on these comets will be described later. The future motion of C/1999~F1
is the most interesting case of all LPCs investigated by us so far.
According to our calculations, it will nominally pass as close as
0.053\,pc (less than 11\,000\,au) from Gliese~710 (see Section
\ref{sec:Examples-of-the} for details).

Since the numerical integration of cometary motion with such a complex
dynamical model is time-consuming, we decided also to check the influence
of each individual star on the motion (past and future) of each individual
object from among 108 LPCs at hand. It is a well-known
fact that stars with small masses and/or high total velocities have
rather small influence on cometary motion, but every singular case
has to be checked because of possible close encounters and high dependence
of the net effect on a stellar passage geometry.

We performed an additional test in which each nominal orbit of 108~comets
were numerically integrated backward and forward as many times as
the number of included stellar perturbers. In each run one star was
``switched off'' so its influence on the cometary motion (a direct
one as well as an indirect through the change of the other stars trajectories)
can be monitored.

As a result we found that 9 stars from the P group, 12 from the F
group and 4 from the C group can be safely removed from our list of
potential stellar perturbers since they change previous/next perihelion
passages of any tested comet by less than 1 per cent. Additionally, three
stars should be reclassified from C group, two of them to the P group
and one to the F group (they act significantly only in the past or
future motion of comets respectively).

After these modifications, we have 4~stars or stellar systems in
group~C, 32 in group~P and 29 in group~F. As a result, we can include
only 36 stellar perturbers (P+C) when integrating the past cometary
motion and only 33 stars (F+C) for the future motion instead of using
all 90~stars for each cometary swarm of thousands of VCs, what significantly
speed-up our calculations.

\section{Past motion of the sample of 22 LPCs discovered in 2006-2010}

\label{sec:Past-motion}

Our main purpose when analysing past motions of LPCs is to search
for the apparent source/sources of these comets. At the first step
we attempt to discriminate between dynamically old and new comets.
As we conclude in our previous papers, one cannot do it only on the
basis of the original semimajor axis (as it is still widely discussed
in many papers) but it seems to be necessary to examine the previous
perihelion distance of each comet. Similar to our Paper~II we decided
to apply three different threshold values for $q_{{\rm {prev}}}$,
namely 10, 15 and 20\,au. In our calculations all comets are replaced
with 5\,001\,VCs, so these thresholds are applied individually to
each returning VC. If $q_{{\rm {prev}}}<10$\,au we call this VC
a dynamically old one. If $q_{{\rm {prev}}}>20$\,au we call it a
new one. Finally, for 10\,au$\leqslant q_{{\rm {prev}}}\leqslant20$\,au
we call it uncertain. Taking all this together, we are rather careful
in drawing conclusions: we call a particular comet the dynamically
new only if all its VCs are dynamically new, and we call a comet the
dynamically old if all its VCs are dynamically old. In the rest of
cases, we treat a comet as having the uncertain dynamical status.

\noindent The results of our past motion investigations are presented
in three tables: Table~\ref{tab:past_motion_old} containing eight
dynamically old comets, Table~\ref{tab:past_motion_new} containing
nine dynamically new comets and Table~\ref{tab:past_motion_uncertain}
showing five uncertain cases. In each table we present two rows of
results for each comet: first one (highlighted by grey shading
consists of the results obtained from calculations which include all
stellar perturbations described earlier. The second row presents,
for comparison purposes, the results obtained when the stellar perturbations
are completely omitted. We present such a detailed comparison to validate
our general opinion that stellar perturbations from all known (so
far) stars do not change past motion of any analysed comets significantly,
especially from the point of view of their dynamical status. Note,
that throughout all these studies, we use a modified original and
future orbit quality classes, described in detail in Part~I of this
investigation \citep{kroli-dyb:2013} as well as recently in \citet{krolikowska:2014}.
For the description of model names see also Part~I.

\begin{figure}
\includegraphics[angle=270,width=1\columnwidth]{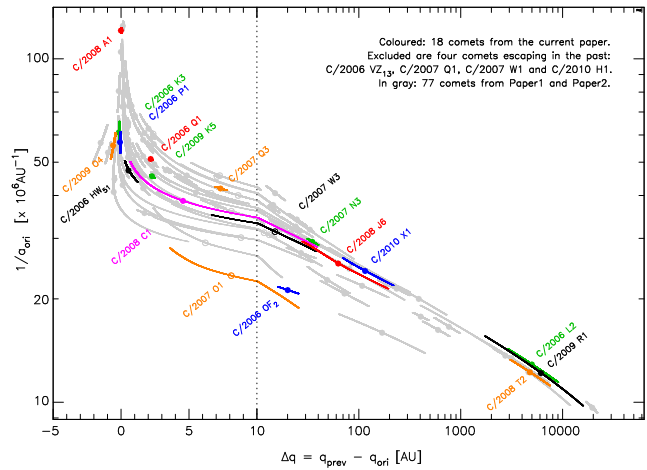}

\caption{\label{fig:Overall-dependence-13}Overall dependence of the perihelion
distance change during the last orbital revolution on the original
semimajor axis. Full dots represent nominal orbits of dynamically
new and old comets, open circles denote uncertain dynamical status.
Accompanying curves represent errors derived from swarms of VCs, truncated
at 1$\sigma$ level. Note that the horizontal scale is logarithmic
right from 10\,au and linear otherwise.}
\end{figure}

All comets from Table~\ref{tab:past_motion_old} have entire swarm
of their VCs elliptical and returning, i.e. their aphelion distances
are smaller than the adopted here escape threshold of 120\,000\,au
for all VCs. Only one comet (C/2007~Q3) has previous perihelion close
to 10\,au, all other have $q_{{\rm {prev}}}<5$\,au. There are four
comets (C/2006~K3, C/2006~P1, C/2008~A1, C/2009~O4) with previous
perihelion distances even smaller than the observed one (see also
Fig.~\ref{fig:Overall-dependence-13}). These four comets have relatively
small semimajor axes. It is rather counter-intuitive: they have very
slow orbital evolution resulting from Galactic perturbations, so they
should visited the inner planetary system several times in the past
(if not tens of times). This fact, together with their small perihelion
distances, make them suspiciously immune from planetary perturbations
and physical ageing.

The largest semimajor axis in dynamically old comets group is about
24\,000\,au, what corresponds to the orbital period of about 4\,Myr.
We are rather convinced that it is really improbable that we missed
an important (i.e. massive and/or slow) stellar perturber which visited
the solar neighbourhood closer than 0.5\,pc in this time interval.
As a result, we can deliberately state that this eight comets are
certainly dynamically old and they visited the inner part of the Solar
system at least once in the past.

Among nine dynamically new comets (Table \ref{tab:past_motion_new})
the situation is more complicated. We have one comet, C/2007~W1,
with all VCs escaping in the past along hyperbolic orbits. This object
is the only serious candidate for an interstellar comet among all
LPCs investigated by us so far. Comet C/2010~H1 has also a negative
$1/a_{{\rm prev}}$-value but its orbit is poorly known. We have other
two comets (C/2007~N3 and C/2010~X1) with a whole swarm of VCs returning
and one comet, C/2008~T2, with all VCs elliptical but escaping, i.e.
crossing our escape threshold of 120\,000\,au. The remaining five
comets have their VCs swarms mixed -- some of VCs are returning and
some are escaping. The smallest semimajor axis in this group is about
34\,000\,au.

The remaining five comets (Table~\ref{tab:past_motion_uncertain})
we have classified as of uncertain dynamical status. Comet C/2006~OF$_{2}$
seems to be dynamically new but 22\,per cent of its VCs have their
previous perihelion distances below 20\,au (and among them 3\,per
cent placed below 15\,au). Due to relatively large semimajor axis
of this comet (about 48\,000\,au) its motion is rather sensitive
even to weak stellar impulses. On the other hand, comet C/2007~Q1
has the orbit of poorest quality in the sample studied here (data-arc
span of only 24\,days), and as a result it has almost 50\,per cent
of hyperbolic VCs as well as the mean previous perihelion as small
as 3\,au for the returning part of its VCs swarm.

\begin{table*}
\caption{\label{tab:past_motion_new} The past distributions of swarms of VCs
for dynamically new comets. For details see the description of Table~\ref{tab:past_motion_old}
`S' suffix denotes results for the swarm synchronously stopped while
`R' suffix informs, that we present the result only for the returning
part of the VCs swarm.}

\begin{tabular}{llccccccc}
\hline 
Comet  & model  & $q_{{\rm osc}}$  & \multicolumn{3}{c}{Number of VCs} & $Q_{{\rm prev}}$  & $q_{{\rm prev}}$  & $1/a_{{\rm prev}}$\tabularnewline
 & \& quality  & au  & {[}R{]}  & {[}E{]}  & {[}H{]}  & $10^{3}$au  & au  & $10^{-6}$au$^{-1}$\tabularnewline
 & class  &  &  &  &  &  &  & \tabularnewline
\hline 
{[}1{]}  & {[}2{]}  & {[}3{]}  & {[}4{]}  & {[}5{]}  & {[}6{]}  & {[}7{]}  & {[}8{]}  & {[}9{]}\tabularnewline
\hline 
\rowcolor{LightGray}C/2006 L2 McNaught  & GR, 1a  & 1.99  & 177  & 4824  & 0  & 128 - 144 - 165S  & 1225 - 1607 - 1822R  & 13.87\textpm{} 1.37S\tabularnewline
 &  &  & 173  & 4828  & 0  & 128 - 143 - 164S  & 1145 - 1574 - 1796R  & 13.93\textpm{} 1.36S\tabularnewline
\rowcolor{LightGray}C/2006 VZ\textsubscript{13} LINEAR  & NG-PRE, 2a  & 1.02  & 1441  & 3560  & 8  & 100 - 143 - 258S  & 83 - 286 - 570R  & 13.91 \textpm{} 4.80S\tabularnewline
 &  &  & 1440  & 3561  & 8  & 100 - 143 - 259S  & 93 - 318 - 623R  & 13.91 \textpm{} 4.80S\tabularnewline
\rowcolor{LightGray}C/2007 N3 Lulin  & GR-PRE, 1a  & 1.21  & 5001  & 0  & 0  & 68.25 \textpm{} 1.37  & 31.2 - 36.3 - 42.3  & 29.30 \textpm{} 0.59\tabularnewline
 &  &  & 5001  & 0  & 0  & 68.27 \textpm{} 1.38  & 30.6 - 35.5 - 41.4  & 29.29 \textpm{} 0.59\tabularnewline
\rowcolor{LightGray}C/2007 W1 Boattini  & NG-PRE, 1b  & 0.85  & 0  & 5001  & 5001  & --  & --  & -42.75 \textpm{} 2.34\tabularnewline
 &  &  & 0  & 5001  & 5001  & \emph{--}  & \emph{--}  & -42.81 \textpm{} 2.35\tabularnewline
\rowcolor{LightGray}C/2008 J6 Hill  & GR, 1b  & 2.00  & 4945  & 56  & 0  & 65.6 - 78.7 - 97.5  & 23.8 - 63.5 - 266R  & 25.50 \textpm{} 3.84\tabularnewline
 &  &  & 4946  & 55  & 0  & 65.6 - 78.5 - 97.4  & 24.3 - 72.5 - 291R  & 25.54 \textpm{} 3.84\tabularnewline
\rowcolor{LightGray}C/2008 T2 Cardinal  & GR, 1b  & 1.20  & 0  & 5001  & 0  & 144 - 158 - 177  & 25.3 \textpm{} 1.5  & 12.62 \textpm{} 1.05\tabularnewline
 &  &  & 0  & 5001  & 0  & 143 - 158 - 177  & 27.3 \textpm{} 1.6  & 12.67 \textpm{} 1.05\tabularnewline
\rowcolor{LightGray}C/2009 R1 McNaught  & NG, 1b  & 0.41  & 778  & 4223  & 0  & 117 - 153 - 223S  & 478 - 1129 - 1614R  & 13.04 \textpm{} 3.20S\tabularnewline
 &  &  & 771  & 4230  & 0  & 117 - 153 - 224S  & 470 - 1110 - 1610R  & 13.02 \textpm{} 3.20S\tabularnewline
\rowcolor{LightGray}C/2010 H1 Garradd  & GR,2b  & 2.75  & 176  & 4826  & 3096  & --  & --  & -3.22 \textpm{} 10.73S\tabularnewline
 &  &  & 176  & 4825  & 3088  & \emph{--}  & \emph{--}  & -3.17 \textpm{} 10.71S\tabularnewline
\rowcolor{LightGray}C/2010 X1 Elenin  & GR\textsubscript{May},1b  & 0.48  & 5001  & 0  & 0  & 74.3-83.0-93.7  & 61.7 - 116 - 260  & 24.07 \textpm{} 2.20\tabularnewline
 &  &  & 5001  & 0  & 0  & 74.3-82.9-93.6  & 65.5 - 132 - 290  & 24.11 \textpm{} 2.20\tabularnewline
\hline 
\end{tabular}
\end{table*}

\begin{table*}
\caption{\label{tab:past_motion_uncertain} The past distributions of swarms
of VCs for comets with the uncertain dynamical status. For details
see the description given in Table~\ref{tab:past_motion_old}. Additionally,
in col.~{[}10{]} we presented the statistics of each VC swarm from
their dynamical status point of view. Suffixes `S' and `R' are used
as in Table~\ref{tab:past_motion_new}.}

\begin{tabular*}{1\textwidth}{@{\extracolsep{\fill}}llcccccccc}
\hline 
Comet  & model \& & $q_{{\rm osc}}$  & \multicolumn{3}{c}{Number of VCs} & $Q_{{\rm prev}}$  & $q_{{\rm prev}}$  & \multicolumn{1}{c}{$1/a_{{\rm prev}}$} & \% of VCs\tabularnewline
 & quality  & au  & {[}R{]}  & {[}E{]}  & {[}H{]}  & $10^{3}$au  & au  & \multicolumn{1}{c}{$10^{-6}$au$^{-1}$} & dyn. new at:\tabularnewline
 & class  &  &  &  &  &  &  &  & 10\,au - 15\,au - 20\,au \tabularnewline
\hline 
{[}1{]}  & {[}2{]}  & {[}3{]}  & {[}4{]}  & {[}5{]}  & {[}6{]}  & {[}7{]}  & {[}8{]}  & {[}9{]}  & {[}10{]} \tabularnewline
\hline 
\rowcolor{LightGray}C/2006 OF\textsubscript{2 }Broughton  & NG, 1a+  & 2.43 & 5001 & 0  & 0  & 94.6 \textpm{} 2.2  & 17.2 - 22.5 - 29.5  & 21.15 \textpm{} 0.49  & 100.0 - 97.4 - 71.5\tabularnewline
 &  &  & 5001 & 0  & 0  & 94.6 \textpm{} 2.2  & 28.3 - 36.3 - 46.6  & 21.15 \textpm{} 0.49  & 100.0 - 100.0 - 99.9\tabularnewline
\rowcolor{LightGray}C/2007 O1 LINEAR  & GR, 1a  & 2.88 & 4612 & 389  & 0  & 67.9 - 86.2 - 114.7S & 5.70 - 10.2 - 27.0R & 23.32 \textpm{} 4.73S  & 54.7 - 36.8 - 25.7\tabularnewline
 &  &  & 4612 & 389  & 0  & 67.9 - 86.2 - 114.8S & 5.54 - 9.71 - 21.0R & 23.32 \textpm{} 4.74S  & 51.9 - 29.7 - 18.6\tabularnewline
\rowcolor{LightGray}C/2007 Q1 Garradd  & GR, 3a  & 3.01 & 2630 & 2371 & 2323 & 1.54 - 3.57 - 17.02R & 2.98 - 3.01 - 3.17R & 47.29 \textpm{} 798.91S  & 49.0 - 48.8 - 48.6\tabularnewline
 &  &  & 2630 & 2371 & 2323 & 1.54 - 3.57 - 17.02R & 2.98 - 3.00 - 3.16R & 47.30 \textpm{} 798.89S  & 49.0 - 48.7 - 48.6\tabularnewline
\rowcolor{LightGray}C/2007 W3 LINEAR  & NG, 1b  & 1.78 & 5001 & 0  & 0  & 55.3 - 64.0 - 75.9 & 7.09 - 17.3 - 50.6 & 31.27 \textpm{} 3.85  & 76.8 - 57.0 - 42.8\tabularnewline
 &  &  & 5001 & 0  & 0  & 55.3 - 64.0 - 75.7 & 6.95 - 17.0 - 52.0 & 31.27 \textpm{} 3.85  & 75.9 - 56.2 - 42.3\tabularnewline
\rowcolor{LightGray}C/2008 C1 Chen-Gao  & GR, 2a  & 1.26 & 4849 & 152  & 2  & 37.3 - 51.6 - 79.3R & 1.64 - 5.69 - 81.1R & 38.47 \textpm{} 11.70S  & 39.6 - 32.8 - 28.7\tabularnewline
 &  &  & 4850 & 151  & 2  & 37.3 - 51.6 - 79.2R & 1.78 - 5.93 - 88.2R & 38.49 \textpm{} 11.70S  & 40.0 - 33.1 - 29.0\tabularnewline
\hline 
\end{tabular*}
\end{table*}

In Fig.~\ref{fig:Overall-dependence-13} we present the overall dependence
of the perihelion distance change during the last orbital revolution
on the original semimajor axis. Coloured plots present results for
comets from the currently studied sample, and in grey we present the
results for 77~comets from Paper~I and II. Each full circle denotes
a nominal orbit of dynamically old or new comet while open circles
mark uncertain cases. Thin lines represent here the uncertainty intervals
and are constructed from our VCs swarms after limiting them to 1$\sigma$
intervals. It is a well-known fact that due to the
dominating role of the Galactic perturbations we observe comets mainly
in the decreasing phase of their perihelion distance evolution ($\Delta q=q_{\rm prev}-q_{\rm orig}>0$).
However, some comets were observed during the increasing part of their
perihelion distance evolution, so we decided to present also negative
$\Delta q$ values in Fig.~\ref{fig:Overall-dependence-13} by means
of switching the horizontal scale from logarithmic to linear one for
$\Delta q<$10\,au. The strong asymmetry of the
plot in Fig.\ref{fig:Overall-dependence-13} with respect to zero
is the evidence, that Galactic perturbations are the dominating 
mechanism of producing observable comets. If the LPCs population was significantly
thermalised by stellar perturbations this plot would be much more
symmetric. Additionally, if the Jupiter-Saturn barrier was completely
opaque the whole region near zero would be empty.

The dominance of the decreasing $q_{{\rm prev}}$ phase is also manifested
in the argument of perihelion of the observed LPCs when expressed
in the Galactic frame, see Fig.~\ref{fig:rozetka-22}. All dynamically
new comets have their previous argument of perihelion in the first
or third quarter, which corresponds to the decreasing phase of the
perihelion distance distribution. The main concept of this figure
is inspired by Fig.~3 from \citet{fouchard-r-f-v:2011}, but instead
of their schematic draw, the real dynamical evolution are plotted
here. This figure is a clear evidence that perturbations from any
known star do not change significantly the past motion of the studied
comets during their last orbital revolution. For a comparison purpose,
we present similar plot for 86~LPCs studied in Paper~I and II, see
Fig.~\ref{fig:rozetka-86}. Again, all dynamically new comets have
their argument of perihelion (in a Galactic frame) in the first or
third quarter. One can also notice, that we did not obtain even one
case in which a significant stellar action can be seen.

\section{Future motion}

\label{sec:Future-motion}

As it regards future motion of 21~LPCs studied in this paper (C/2010~X1
disintegrated during the observed perihelion passage) the situation
is quite typical and completely different from the past orbital evolution.
Generally, we obtain two types of motion. Six comets will leave the
Solar system along hyperbolic orbits (Table~\ref{tab:future_motion_escaping})
as a result of planetary perturbations. The same planetary action
resulted in significant shortening of the semimajor axes of remaining
15~comets studied here (Table \ref{tab:future_motion_returning}),
with the extreme case of C/2007~N3, for which $a_{{\rm fut}}\approx1\,215$\,au,
obtained from $a_{{\rm orig}}\approx$34\,000\,au.

Only two comets in the later group have mixed VCs swarms. The extremely
dispersed VCs swarm of C/2007~Q1 (as a result of largest orbital
uncertainties) consists of 1\,306 returning clones and 3\,695 escaping
ones (3\,657 of them are hyperbolic). The mixed swarm of C/2009~R1
consists mainly of returning VCs but about 20~per cent of clones
are escaping (most often along hyperbolic orbits). This swarm is highly
dispersed since future orbit of this comet is rather uncertain --
all observations were obtained before its perihelion passage, for
details see Part I. The remaining 13~comets from Table~\ref{tab:future_motion_returning}
have all their VCs returning with a typical perihelion distance of
about 1\,au or even less.

\begin{figure}
\begin{centering}
\includegraphics[clip,width=0.8\columnwidth]{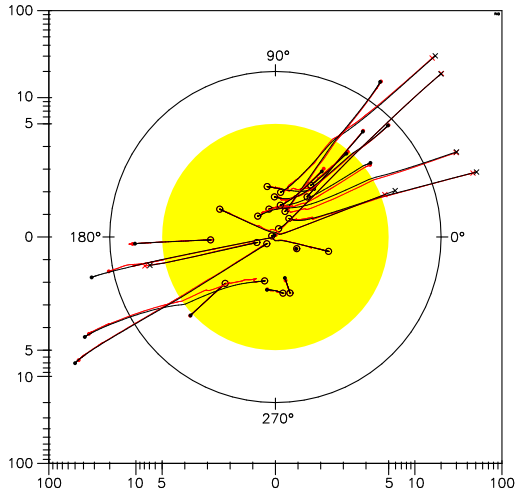} 
\par\end{centering}

\caption{\label{fig:rozetka-22}Last orbital period evolution of the osculating
argument of perihelion and the perihelion distance of nominal orbits
of 22 comets investigated in this paper. For each comet a plot in
polar coordinates is presented, in which the distance from the centre
equals the osculating perihelion distance while the ``phase angle''
equals the osculating argument of perihelion. Open circles denote
start of each calculation with the original orbit, full dot marks
a previous perihelion. For escaping comets we stopped the calculation
at 120\,000 au from the Sun and this fact is marked with a cross.
Two different calculations are presented for each comet: red one with
all stellar perturbations included, overprinted with the black plot
where only Galactic perturbations are taken into account. In the yellow
area ( $q<5$~au, the observability zone) the scale of the radial
coordinate is linear while outside this area it is logarithmic.}
\end{figure}

\begin{figure}
\begin{centering}
\includegraphics[clip,angle=270,width=0.9\columnwidth]{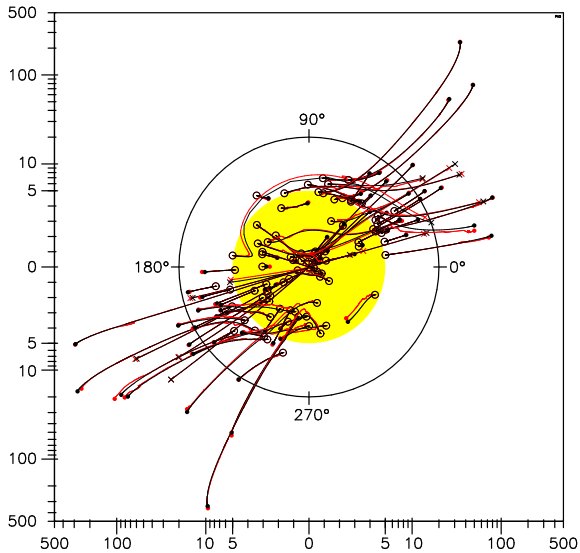} 
\par\end{centering}

\caption{\label{fig:rozetka-86}The same as in Fig.\ref{fig:rozetka-22}, but
for 86~LPCs studied in Paper I and II (note that a radial coordinate
scale has to be changed to show all these comets in one figure).}
\end{figure}

\begin{figure}
\begin{centering}
\includegraphics[angle=270,width=0.8\columnwidth]{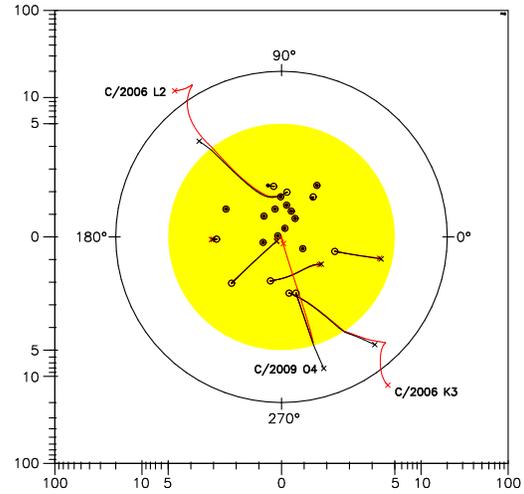} 
\par\end{centering}

\caption{\label{fig:rozetka-fut-21}The same as in Fig.\ref{fig:rozetka-22}
but for future motion (only 21 comets are here since C/2010~X1 disintegrated
during the observed perihelion passage).}
\end{figure}

\begin{center}
\begin{table}
\caption{\label{tab:future_motion_escaping}The $1/{\rm a}_{{\rm next}}$ values
for six comets with all 5001 VCs escaping in the future in hyperbolic
orbits. We present two rows for each comet: the upper row (highlighted
by grey shading) contains the results with the stellar perturbations
included, and the second row (presented here for comparison) shows
results where stellar perturbations were ignored in the dynamical
calculations.}

\centering{}%
\begin{tabular}{ccc}
\hline 
Comet  & model \&  & $1/a_{{\rm next}}$\tabularnewline
 & quality class  & $10^{-6}$au$^{-1}$\tabularnewline
\hline 
{[}1{]}  & {[}2{]}  & {[}3{]}\tabularnewline
\hline 
\rowcolor{LightGray}C/2006 K3  & NG, 1a  & -130.27 \textpm{} 4.64\tabularnewline
 &  & -130.26 \textpm{} 4.65\tabularnewline
\rowcolor{LightGray}C/2006 L2  & GR, 1a  & -93.97 \textpm{} 1.41\tabularnewline
 &  & -93.96 \textpm{} 1.41\tabularnewline
\rowcolor{LightGray}C/2006 OF\textsubscript{2}  & NG, 1a+  & -658.76 \textpm{} 0.23\tabularnewline
 &  & -658.76 \textpm{} 0.23\tabularnewline
\rowcolor{LightGray}C/2007 O1  & GR, 1a  & -496.94 \textpm{} 4.69\tabularnewline
 &  & -496.99 \textpm{} 4.69\tabularnewline
\rowcolor{LightGray}C/2008 J6  & GR, 1b  & -479.37 \textpm{} 3.99\tabularnewline
 &  & -479.35 \textpm{} 3.99\tabularnewline
\rowcolor{LightGray}C/2009 O4  & GR, 1b  & -57.2 1\textpm{} 4.82\tabularnewline
 &  & -57.89 \textpm{} 4.83\tabularnewline
\hline 
\end{tabular}
\end{table}

\par\end{center}

There is a significant difference between past and future motion in
the context of the importance of stellar perturbations. As it was
clearly presented in Figs.~\ref{fig:rozetka-22} and \ref{fig:rozetka-86}
stellar perturbations do not change the past dynamical cometary evolution
in a significant manner during their last orbital period. As it concerns
their future motion situation might be completely different since
we expect at least one very close stellar encounter with the Sun.
The star Gliese~710 (HIP~89825) will pass as close as about 0.24\,pc
from the Sun in 1.4\,Myr. As a result, arbitrarily small comet-star
distance may occur, provided  the orbital period is
longer than 1.4\,Myr (we restricted our calculations to one orbital
period to the past and in future). None of the comets from Table~\ref{tab:future_motion_returning}
has orbital period long enough to meet Gliese~710, but all comets
from Table~\ref{tab:future_motion_escaping} suffer from weaker or
stronger gravitational kick from this star, as it is shown in Fig.~\ref{fig:rozetka-fut-21}
and  will be discussed in detail in Section \ref{C/2009-O4} on the past dynamical evolution of C/2009~O4. However,
even the strongest effects, visible in cases of C/2006~K3 and C/2006~L2,
do not change their fate -- they will still leave our planetary system
along (only slightly modified) hyperbolic orbits.

\begin{table*}
\caption{\label{tab:future_motion_returning}The future elements distributions
of 16 comets with returning or mixed swarms of VCs in terms of returning
{[}R{]}, escaping {[}E{]}, including hyperbolic {[}H{]} VC numbers.
We present two rows for each comet: the upper row (highlighted by
grey shading) contains the results with the stellar perturbations
included and the second row (presented here for comparison) shows
results where stellar perturbations were ignored in our dynamical
calculations. Next aphelion (col.~6) and perihelion (col.~7) distances
are described either by a mean value of the approximate normal distributions,
or three deciles at 10, 50 (i.e. median), and 90 per cent in cases
where the normal distribution is not applicable. In the case of mixed
swarm the mean values or decides of $Q$ and $q$ are given for the
returning (suffix `R') part of the VCs swarm, where the escape limit
of 120\,000\,au was generally adopted. Last column presents the
value of $1/{\rm a}_{{\rm next}}$. Suffix `S' denotes a result for
the swarm synchronously stopped.}

\begin{tabular}{llcccccc}
\hline 
Comet  & model  & \multicolumn{3}{c}{Number of VCs} & $Q_{{\rm next}}$  & $q_{{\rm next}}$  & $1/a_{{\rm next}}$\tabularnewline
 & \& quality  &  &  &  & $10^{3}$au  & au  & $10^{-6}$au$^{-1}$\tabularnewline
 & class  & {[}R{]}  & {[}E{]}  & {[}H{]}  &  &  & \tabularnewline
\hline 
{[}1{]}  & {[}2{]}  & {[}3{]}  & {[}4{]}  & {[}5{]}  & {[}6{]}  & {[}7{]}  & {[}8{]}\tabularnewline
\hline 
\rowcolor{LightGray}C/2006 HW\textsubscript{51}  & GR, 1a  & 5001  & 0  & 0  & 22.25\textpm{}0.84  & 2.33 - 2.35 - 2.37  & 90.02\textpm{}3.37 \tabularnewline
 &  & 5001  & 0  & 0  & 22.25\textpm{}0.84  & 2.34 - 2.35 - 2.37  & 90.01\textpm{}3.37 \tabularnewline
\rowcolor{LightGray}C/2006 P1  & NG, 1b  & 5001  & 0  & 0  & 4.28\textpm{}0.03  & 0.172140\textpm{}0.000005  & 467.64\textpm{} 3.56\tabularnewline
 &  & 5001  & 0  & 0  & 4.28\textpm{}0.03  & 0.172225\textpm{}0.000002  & 467.64\textpm{} 3.56\tabularnewline
\rowcolor{LightGray}C/2006 Q1  & NG, 1a+  & 5001  & 0  & 0  & 2.824\textpm{}0.001  & 2.759058\textpm{}0.000001  & 707.45\textpm{} 0.31\tabularnewline
 &  & 5001  & 0  & 0  & 2.824\textpm{}0.001  & 2.759930\textpm{}0.000001  & 707.44\textpm{}0.31 \tabularnewline
\rowcolor{LightGray}C/2006 VZ\textsubscript{13}  & NG, 1b  & 5001  & 0  & 0  & 4.07\textpm{}0.17  & 1.0121\textpm{}0.0001  & 491.6\textpm{} 20.1\tabularnewline
 &  & 5001  & 0  & 0  & 4.07\textpm{}0.17  & 1.0121\textpm{}0.00001  & 491.6\textpm{} 20.1\tabularnewline
\rowcolor{LightGray}C/2007 N3  & NG-POST, 1a  & 5001  & 0  & 0  & 2.427\textpm{}0.006  & 1.215060\textpm{}0.000004  & 823.60\textpm{}2.06 \tabularnewline
 &  & 5001  & 0  & 0  & 2.427\textpm{}0.006  & 1.214996\textpm{}0.000005  & 823.60\textpm{}2.06 \tabularnewline
\rowcolor{LightGray}C/2007 Q1  & GR, 3a  & 1306  & 3695  & 3657  & 2.16 - 5.61 - 26.2R  & 2.75-2.97-2.98R  & -463\textpm{}741S \tabularnewline
 &  & 1305  & 3696  & 3657  & 2.16 - 5.61 - 26.2R  & 2.70-2.97-2.99R  & -463\textpm{}741S \tabularnewline
\rowcolor{LightGray}C/2007 Q3  & GR-POST, 1a  & 5001  & 0  & 0  & 15.18\textpm{}0.42  & 2.171\textpm{}0.008  & 131.83\textpm{}3.63 \tabularnewline
 &  & 5001  & 0  & 0  & 15.18\textpm{}0.42  & 2.174\textpm{}0.008  & 131.84\textpm{}3.63 \tabularnewline
\rowcolor{LightGray}C/2007 W1  & NG-POST, 2a  & 5001  & 0  & 0  & 3.61\textpm{}0.05  & 0.84667\textpm{}0.00002  & 554.59\textpm{} 7.09\tabularnewline
 &  & 5001  & 0  & 0  & 3.61\textpm{}0.05  & 0.84633\textpm{}0.00001  & 554.59\textpm{} 7.09\tabularnewline
\rowcolor{LightGray}C/2007 W3  & NG, 1b  & 5001  & 0  & 0  & 5.82\textpm{}0.31  & 1.7700-1.7701-1.7702  & 344.38\textpm{}18.10 \tabularnewline
 &  & 5001  & 0  & 0  & 5.82\textpm{}0.31  & 1.7702-1.7703-1.7704  & 344.37\textpm{} 18.10\tabularnewline
\rowcolor{LightGray}C/2008 A1  & NG-POST, 1b  & 5001  & 0  & 0  & 8.11\textpm{}0.09  & 1.07192\textpm{}0.00003  & 246.52\textpm{}2.82 \tabularnewline
 &  & 5001  & 0  & 0  & 8.11\textpm{}0.09  & 1.07062\textpm{}0.00005  & 246.52\textpm{}2.82 \tabularnewline
\rowcolor{LightGray}C/2008 C1  & GR, 2a  & 5001  & 0  & 0  & 3.98\textpm{}0.09  & 1.26428\textpm{}0.00008  & 502.39\textpm{}11.77 \tabularnewline
 &  & 5001  & 0  & 0  & 3.98\textpm{}0.09  & 1.2627(79)-(82)-(84)  & 502.38\textpm{}11.77 \tabularnewline
\rowcolor{LightGray}C/2008 T2  & GR, 1b  & 5001  & 0  & 0  & 7.25\textpm{}0.03  & 1.20240\textpm{}0.00002  & 275.91\textpm{}1.06 \tabularnewline
 &  & 5001  & 0  & 0  & 7.25\textpm{}0.03  & 1.20195\textpm{}0.00001  & 275.92\textpm{}1.06 \tabularnewline
\rowcolor{LightGray}C/2009 K5  & GR-POST, 1a  & 5001  & 0  & 0  & 3.616\textpm{}0.003  & 1.419967\textpm{}0.000001  & 552.92\textpm{} 0.41\tabularnewline
 &  & 5001  & 0  & 0  & 3.616\textpm{}0.003  & 1.419818\textpm{}0.000001  & 552.92\textpm{}0.41 \tabularnewline
\rowcolor{LightGray}C/2009 R1  & NG, 1b  & 3909  & 1092  & 986  & 4.51 - 9.05 - 32.1R  & 0.386-0.406-0.502R  & 166.3\textpm{} 197.1S\tabularnewline
 &  & 3910  & 1091  & 986  & 4.51 - 9.06 - 32.1R  & 0.390-0.406-0.446R  & 166.3\textpm{}197.1S \tabularnewline
\rowcolor{LightGray}C/2010 H1  & GR, 2b  & 5001  & 0  & 0  & 3.70\textpm{}0.07  & 2.74040\textpm{}0.00006  & 541.02\textpm{}10.77 \tabularnewline
 &  & 5001  & 0  & 0  & 3.70\textpm{}0.07  & 2.7415\textpm{}0.0001  & 541.02\textpm{} 10.77\tabularnewline
\hline 
\end{tabular}
\end{table*}

All significant stellar perturbations visible in Fig.~\ref{fig:rozetka-fut-21}
are caused by Gliese~710. We performed additional test calculation
with this star excluded and all effects of any stellar action almost
disappeared making the character of this plot in this respect similar
to that of Fig.~\ref{fig:rozetka-22}.

\section{Examples of stellar perturbations on LPCs}

\label{sec:Examples-of-the}

In this section we present examples of dynamical evolution of cometary
nominal orbits under simultaneous gravitational action of the whole
Galaxy and individual potential stellar perturbers. During the past
and future comet motion it is possible to observe slow orbital elements
evolution caused by Galactic tides overlapped by (usually very small)
local variations due to the action of passing stars. In many cases,
we identified a star or stellar system responsible for the individual
orbital element fluctuation.

\begin{figure}
\includegraphics[angle=270,width=1\columnwidth]{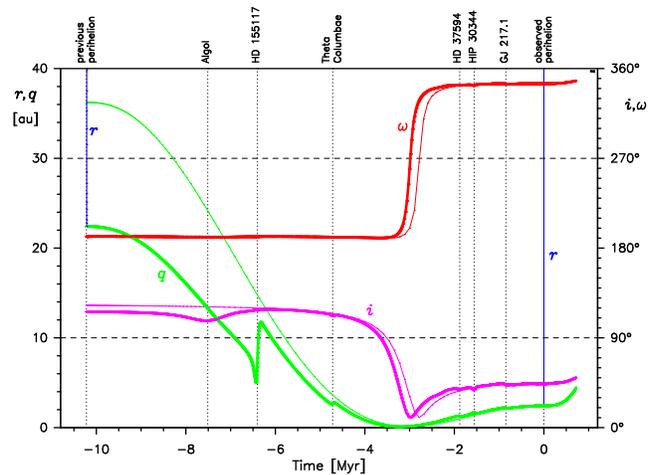}

\caption{\label{fig:C/2006of}Past and future evolution of C/2006~OF$_{2}$
nominal orbit under the simultaneous Galactic and stellar perturbations.}
\end{figure}

\begin{figure}
\includegraphics[angle=270,width=1\columnwidth]{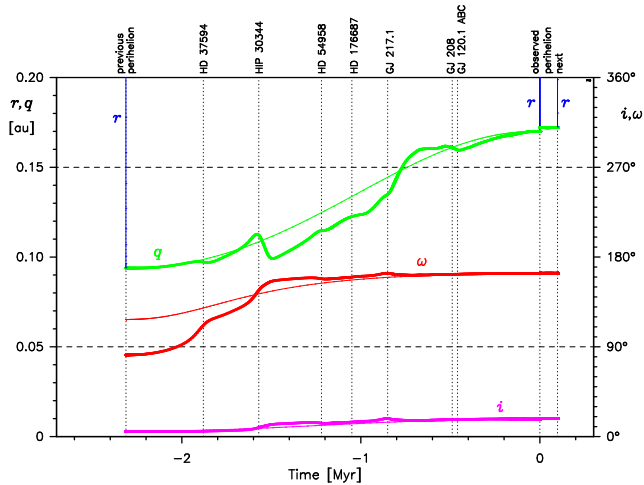}

\caption{\label{fig:C/2006p1}Past and future evolution of C/2006~P1 nominal
orbit under the simultaneous Galactic and stellar perturbations.}
\end{figure}

Figures~\ref{fig:C/2006of}--\ref{fig:C/1999f1} show several examples
of the dynamical evolution of orbits of LPCs, four of them are comets
discovered in the years 2006--2010. In each picture, the horizontal
time axis extends from the previous perihelion passage through the
observed perihelion up to the next perihelion passage or to the moment
of escape (understood here as crossing the threshold heliocentric
distance of 120\,000\,au). The left vertical axis is expressed in
au and corresponds to the osculating perihelion distance plot ($q$,
green line) as well as the heliocentric distance plot ($r$, thin
blue lines). Due to the scales of this pictures the heliocentric distance
plot takes the form of vertical blue lines exactly at perihelion passage
moments. The right vertical axis is expressed in degrees and describes
the evolution of the osculating inclination ($i$, magenta line) and
the argument of perihelion ($\omega$, red line). Both these angular
elements are expressed in the Galactic frame.

\noindent The thick lines depict dynamical evolution under the simultaneous
stellar and Galactic perturbations while the thin lines mark the evolution
with the stellar perturbations excluded. Horizontal dashed lines call
attention to the beginning of the second and fourth quarter of $\omega$,
which values (90\textdegree{} and 270\textdegree{}) are important
from the point of view of the Galactic perturbations.

\noindent The vertical dashed lines show closest approaches of a comet
with the star or stellar system, which name is placed at the top of
the picture. It is worth to mention, that the timing of stellar perturbation
is not necessarily strictly aligned with this closest approach moment
-- it strongly depends on the geometry since the final, heliocentric
orbit change is a net effect of a stellar gravitational action on
both, a comet and the Sun.

\subsection{Comets discovered in the years 2006--2010}

Four characteristic examples how known stellar perturbation can affect
the dynamical evolution of actual near-parabolic comets are illustrated
in Figures~\ref{fig:C/2006of}--\ref{fig:C/2009o4}.

\subsubsection{C/2006~OF$_{2}$ -- comet with the largest change of perihelion
distance in the past due to stellar action}

The largest change in a previous perihelion distance caused by stellar
perturbations can be observed in the case of comet C/2006~OF$_{2}$.
The cumulative perturbations from all stars change the previous perihelion
distance of this comet from 36.3\,au down to 22.5\,au (these are
medians of a non-Gaussian VCs distributions, compare also with nominal
orbital elements given in Table~\ref{tab:past_motion_uncertain}).
Large orbital period (over 10 Myr) have amplified this effect. In
Fig.~\ref{fig:C/2006of} one can easily note a prominent perturbation
by HD~155117 (HIP~84263, 1.3\,M$_{\odot}$, minimal distance from
C/2006~OF$_{2}$ of 0.56\,pc happened 6.4\,Myr ago). Algol system
(HIP~14576, 5.8\,M$_{\odot}$, minimal distance from C/2006~OF$_{2}$
of 2.72\,pc occurred 7.5\,Myr ago) causes only a moderate variation
of the comet inclination. A small local variations of comet inclination
and perihelion distance result from the passage of Theta Columbae
(HIP~29034, 4.14\,M$_{\odot}$, minimal distance from C/2006~OF$_{2}$
of 1.96\,pc occurred 4.7\,Myr ago). No significant change in the
argument of perihelion resulting from the action of these three massive
stars is observed. On the contrary, the cumulative (but weak) perturbations
from HD~37594, HIP~30344 and GJ~217.1 change slightly all three
elements plotted in Fig.\ref{fig:C/2006of}.

\noindent In its future motion, C/2006~OF$_{2}$ do not suffer any
noticeable stellar perturbations and escape from the Solar system
along the strongly hyperbolic orbit (evolution is stopped when a comet
reaches the distance of 120\,000\,au from the Sun).

\subsubsection{C/2006~P1 -- example of cumulative action of stellar perturbation}

Comet C/2006~P1 was observed at a very low perihelion distance of
0.17\,au. According to our calculations its previous perihelion distance
was even much smaller, with a median value of about 0.1\,au (see
Table~\ref{tab:past_motion_old}). It can be seen in Fig.~\ref{fig:C/2006p1}
that the cumulative action of stars on the comet's dynamics does not
change the previous perihelion distance and inclination values noticeably,
while its previous argument of perihelion is changed by over 35 degrees
(from 117\textdegree{} down to 82\textdegree{}). Several individual
weak stellar perturbations are marked in Fig.~\ref{fig:C/2006p1}
but the overall evolution of orbital elements is also affected by
a cumulative effect of other numerous stellar perturbers.

\noindent The narrow range of the left vertical scale of the picture
(0--0.20\,au) allows to show the discontinuity between original and
future orbit (caused by planetary perturbations during the observed
perihelion passage), especially in the perihelion distance. As it
is clearly depicted, planetary perturbations have shortened the semimajor
axis of this comet considerably. Thus, this comet will appear in the
next perihelion passage in a relatively short time of order of 100\,thousand
years having the similarly small perihelion distance.

\noindent It is worth noting, that comet C/2006~P1 McNaught is a
very interesting object. It was very bright with a spectacular tail
spread over half of the sky. Surprisingly, it exhibited moderated
nongravitational effects in the sense of deviation from the pure gravitational
motion, and the nucleus activity became very low soon after perihelion.
This probably can explain how it can survive even several such a close
perihelion passages.

\begin{figure}
\includegraphics[angle=270,width=1\columnwidth]{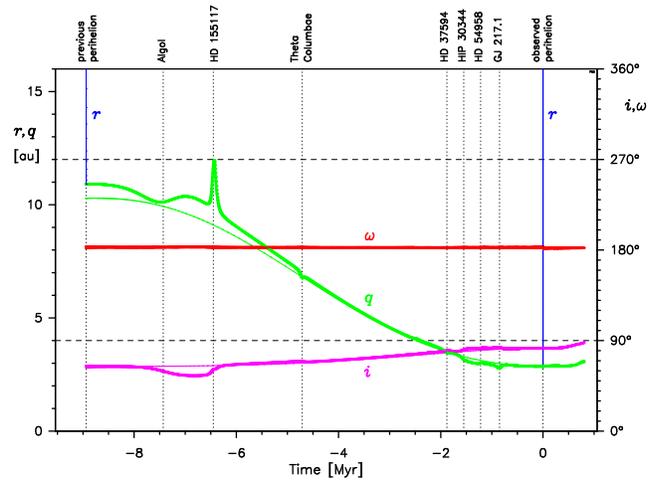}

\caption{\label{fig:C/2007o1}Past and future evolution of C/2007~O1 nominal
orbit under the simultaneous Galactic and stellar perturbations.}
\end{figure}

\subsubsection{C/2007~O1 -- example of a series of weak stellar perturbation}

The case of comet C/2007~O1 is a good example of a very common phenomenon,
when a series of weak stellar perturbations compensate each other.
This is a nature of a single stellar gravitational action on a comet
that the change in osculating elements gained during the star approaching
phase is sometimes completely reversed during its moving away. When
a series of different stellar perturbation occur such a mutual compensation
is a typical feature. In Fig.~\ref{fig:C/2007o1} one can easily
notice, that the net effect of weak perihelion distance (and inclination)
perturbations caused by HD~37594, HIP~30344, GJ~217.1 and HD~54958
is invisible in the plot. The same situation can be observed in a
case of inclination perturbations by Algol system, fully compensated
by the action of HD~155117. However, the perihelion distance perturbations
by these two massive stars do not completely compensate each other.
When following the past motion of C/2007~O1 we obtained its previous
perihelion distance larger by approximately 0.5\,au (the swarm of
VCs is rather highly spread for this comet) after including stellar
perturbations in the dynamical model. It is worth to note that in
this case no changes in osculating argument of perihelion are observed.

\subsubsection{C/2009~O4 -- comet with largest change of perihelion distance due to GJ 710 in the basic sample of 22~comets}

\label{C/2009-O4}

\begin{figure}
\includegraphics[angle=270,width=1\columnwidth]{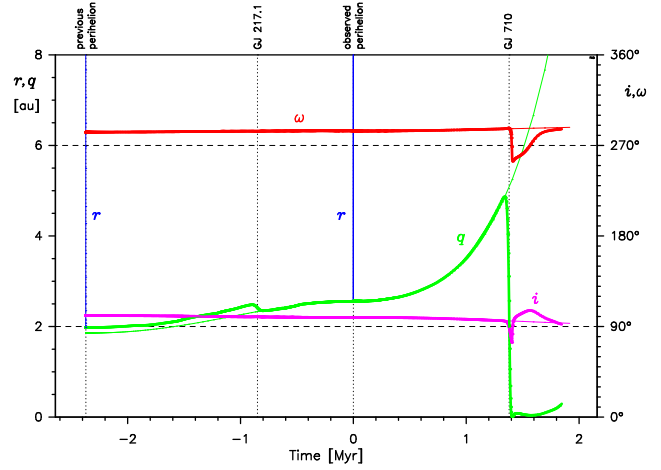}
\caption{\label{fig:C/2009o4}Past and future evolution of C/2009~O4 nominal
orbit under the simultaneous Galactic and stellar perturbations.}
\end{figure}

As it was mentioned earlier, the stellar perturbations along a comet
future motion orbital branch can be much more spectacular. This results
from a future close approach of the star GJ~710 (HIP~89825) and
can potentially lead to an arbitrarily large change in cometary orbits.
However, among 22~LPCs studied in detail in this paper, we did not
found such a drastic case. The largest difference in the future comet
motion we observe in the case of C/2009~O4. This comet escapes from
our planetary system along a hyperbolic orbit ($1/a_{\textrm{fut}}=-56\pm5$\,au$^{-1}$).
In 1.38\,Myr, it will make a moderately close approach to GJ~710
(at a distance of 0.45\,pc, approximately 93\,000\,au) and significant
changes in its osculating elements can be observed, see Fig.~\ref{fig:C/2009o4}.
This is an escaping comet, so, we stopped calculations of its future
motion at a threshold heliocentric distance of 120\,000\,au. The
angular elements, after significant local ``jumps'' remain almost
unchanged at the end of the calculation. It should be noted that the
character of the future motion of C/2009~O4 remains also unchanged
and it will escape along the hyperbolic orbit with almost the same
energy.

\begin{figure}
\includegraphics[angle=270,width=1\columnwidth]{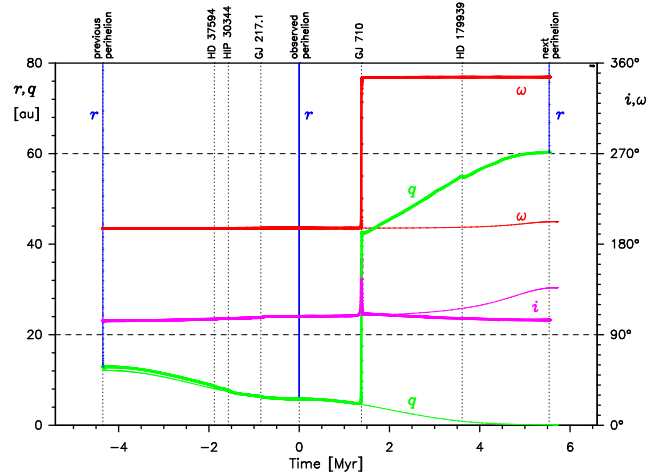}
\caption{\label{fig:C/1999f1}Past and future evolution of C/1999~F1 nominal
orbit under the simultaneous Galactic and stellar perturbations.}
\end{figure}

\subsection{The additional sample of 86~near-parabolic comets}

As stated earlier we have incorporated stellar perturbations in our
standard dynamical model used in long term cometary dynamics investigations.
As a result we decided to check all 86~LPCs previously investigated
by us for any significant changes in orbital elements that result
from the stellar gravitational action.

As it was described in section \ref{sec:Stellar-perturbations-characteristics}
we have not observed any strong stellar perturbations in the past
motion of these comets but their future motion exhibits several interesting
cases. As this paper is mainly devoted to LPCs discovered in 2006-2010
we present only one extreme example.

\subsubsection{C/1999~F1 -- the extreme case of the GJ~710 action due to very
close encounter}

We found C/1999~F1 to be a comet which will make the closest approach
to GJ~710. Following the nominal orbit of this comet we found the
minimal comet-star distance to be as small as 11\,000\,au. Despite
the moderate mass of GJ~710 such a close approach will cause large
changes in future C/1999~F1 orbit, as it is depicted in Fig.\ref{fig:C/1999f1}.

In contrary to C/2009~O4 case, here we note large and persistent
change in the osculating argument of perihelion of C/1999~F1, that
causes a switch of the phase of the Galactic dynamical evolution,
from decreasing to increasing of the perihelion distance. In Paper~II
(where only Galactic perturbations were taken into account) we reported
that the next perihelion distance of this comet will be as small as
0.11\,au. Now, after allowing for the stellar perturbations one can
clearly see in Fig.~\ref{fig:C/1999f1} that its next perihelion
distance will be over 60\,au! A small contribution to that drastic
change will also be given by HD~179939 (HIP~94512, mass 2.5\,M$_{\odot}$,
minimal distance to C/1999~F1 will go down to 1.78\,pc in 3.6\,Myr).
Changes in angular elements are also large in the motion of this comet.
The argument of perihelion will change from 202\textdegree{} up to
346\textdegree{} . This crossing of the quarters border at 270\textdegree{}
causes the switch in the Galactic dynamical evolution phase. Osculating
inclination of C/1999~F1 is also significantly influenced by GJ~710,
decreasing from 137\textdegree{} down to 105\textdegree{}.

Such a large orbit change caused by GJ~710 remains in drastic contrast
to barely noticeable results of cumulative stellar perturbations during
the C/1999~F1 past orbital revolution.

\section{Dynamically new versus old near-parabolic comets}

\label{sec:dynamically_new_old_}

\vspace{0.2cm}

\begin{figure}
\begin{centering}
\includegraphics[width=1\columnwidth]{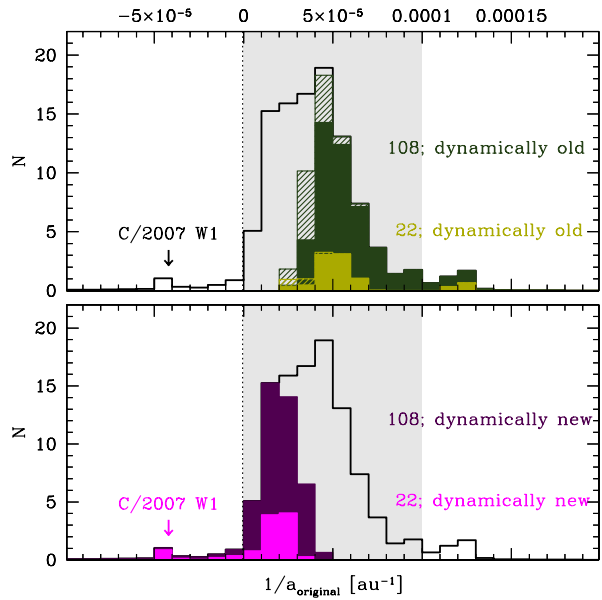} 
\par\end{centering}

\caption{\label{fig:stat108_new_old_uncertain}Upper panel: dark green/gold
histograms show the distribution of dynamically old VCs for the samples
of 108/22~comets, where the dashed parts represent the dynamically
uncertain VCs. Lower panel: Histograms filled in violet/magenta ink
present the distributions for the dynamically new VCs for the same
samples. The overall distribution of the entire sample of 108~comets
is given by histogram shown in black ink in both panels.}
\end{figure}

In this section, we discuss the dynamical status of 22~comets analysed
here in details in comparison to the entire sample of 108~comets
investigated by us in last few years. However, for nine comets considered
in Papers~I--III (C/1990~K1, C/1993~A1, C/1997~A1, C/1999~F1,
C/1999~N4, C/2002~J4, C/2003~K4, C/2003~S3 and C/2005~K1), we
updated the orbital solutions by taking the newer osculating orbits
described as \emph{new solution} in Table~A.1 of \citet{krolikowska:2014}
for the dynamical evolution. As a result, the overall histogram of
$1/a_{{\rm orig}}$ shown in thick black ink in Fig.~\ref{fig:stat108_new_old_uncertain}
is slightly different from that presented in Fig.~9 in \citet{kroli-dyb:2013}.
Additionally, we repeated all our previous dynamical calculations
using a model with stellar perturbations described in Section~\ref{sec:Stellar-perturbations}.
However, we should emphasise that in a statistical sense, stellar
perturbations have not changed any of histograms presented in Fig.~\ref{fig:stat108_new_old_uncertain}.

For constructing the histograms, we treated each VC from the swarm
individually. This means that overall distribution given in Fig.~\ref{fig:stat108_new_old_uncertain}
is composed of the 108~individual and normalised $1/a_{{\rm orig}}$-distributions
(each based on 5\,001\,VCs). Additionally, to prepare the statistics
of dynamically old/new/uncertain, we also considered each VC individually:
a VC was defined as a `dynamically old' when its $q_{{\rm prev}}$
was smaller than 10\,au and $Q_{{\rm prev}}$ was inside a sphere
of 120\,000\,au, the dynamical status was called `uncertain' for
10\,au\,$\le q_{{\rm prev}}<$\,20\,au and $Q_{{\rm prev}}<$120\,000\,au,
the remaining VCs were dynamically new. The effect of this approach
can be seen in Fig.~\ref{fig:stat108_new_old_uncertain}, where the
coloured histograms in the upper panel show the distributions of dynamically
old part of distributions and dashed parts of bars represent the uncertain
VCs, whereas the colour histograms in the lower panel highlight dynamically
new parts of the entire distributions.

The distributions of dynamically new and dynamically old comets of
the sample of 22~comets seems to be statistically similar to the
respective distributions representing the entire sample of 108~comets,
with the exception of the occurrence of one definitely hyperbolic
previous orbit of C/2007~W1.

The percentage contribution of dynamically new/old/uncertain to the
respective bins in the range 0\,$\le1/a_{{\rm orig}}<$\,0.000100\,au$^{-1}$
is given in Table~\ref{tab:previous_motion_dynamical_status}. One
can see that in the range of 0.000030\,au$^{-1}\,\le1/a_{{\rm orig}}<$\,0.000040\,au$^{-1}$,
we have less than 60~per cent of the dynamically new comets, and
up to 25~per cent of comets certainly being dynamically old. Even
in the range of 0.000020\,au$^{-1}\,\le1/a_{{\rm orig}}<$\,0.000030\,au$^{-1}$
(33\,000\,au $<a_{{\rm orig}}\leq50\,000$\,au), we are not sure
about the dynamic status of comets, because, qualitatively speaking,
every twenty--thirtieth can be dynamically old comet. This fully confirms
conclusions made earlier with simpler models \citep{dyb-hist:2001,dyb-hab3:2006}.
It seems worth to note that many authors still erroneously  treat LPCs from 
all rows of Table~\ref{tab:previous_motion_dynamical_status} as visiting the 
inner planetary system for the first time, which is not necessarily true.

\begin{figure}
\begin{centering}
\includegraphics[angle=270,width=1\columnwidth]{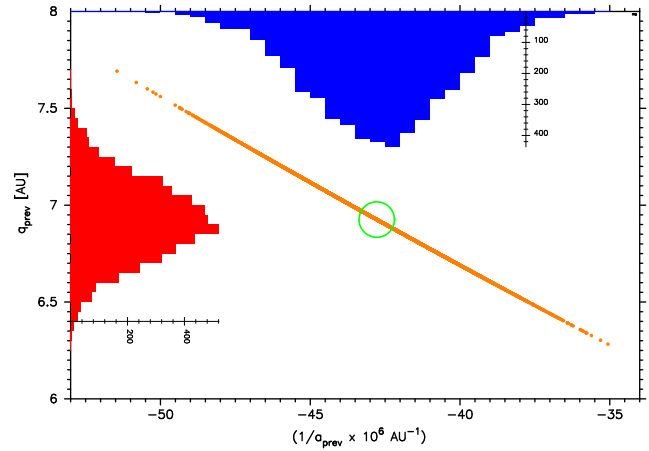} 
\par\end{centering}

\caption{\label{fig:C/2007W1prev}Joint and marginal distributions
of the osculating $1/a$ and $q$ of C/2007~W1 at a heliocentric
distance of 120\,000\,au before entering our planetary system. The
centre of the green circle points to the nominal values. }
\end{figure}

Finally, it is worth to pay attention to these comets, which according
to our investigation come to the Solar system on hyperbolic orbits.
It turns out that we have 3.3~per cent of such VCs in a sample of
108~cometary swarms. It would be formally translated to 3.6~objects,
where 1.8~objects belong to sample of 22~comets analysed here. In
fact, we have, however, only one comet, which is a good candidate
for interstellar comet. This is C/2007~W1~Boattini, widely discussed
in Part~I of these studies (see also Fig.~\ref{fig:stat108_new_old_uncertain}).
Now, after calculating its past motion under the influence of Galactic
tides and all known potential stellar perturbers, we can confirm that
this comet seems to be of interstellar origin. Its osculating perihelion
distance and the inverse semimajor axis distributions are presented
in Fig.\ref{fig:C/2007W1prev} for the heliocentric distance of about
120\,000\,au before the observed perihelion passage (i.e. almost
2~Myr ago). It is worth to mention that the presented $1/a$-distribution
is almost identical to the distribution of the original orbit (at
250\,au from the Sun) and it is not changed by any means by stellar
perturbations. The mean of the $1/a_{{\rm prev}}$-distribution for
this comet is very close to the nominal value and equals (-42.75$\pm$2.34)\,$\times10^{-6}$\,au$^{-1}$.
 It should be stressed that using all available data we obtained hyperbolic 
original orbit for this comet in all investigated models, taking into account 
all uncertainties, also those resulting from strange nongravitational forces 
behaviour. Remembering that this comet reaches the 120000 au escape limit 
in about 2 million years (going backward) it seems really improbable that 
we missed an important stellar perturber in this case. But if  such a perturber 
will be discovered in future, our conclusion might appear inappropriate.

Comet C/2010~H1~Garradd have $1/a_{{\rm orig}}=-3.17\pm10.71$\,au$^{-1}$
in units of $10^{-6}$\,au$^{-1}$, and, in our opinion should be
rather classified as related to Solar system though interstellar origin
can not be ruled out (see also Table~\ref{tab:past_motion_new}).
In the sample of the remaining 86 comets, we did not record similar
cases to C/2007~W1; there were two cases (C/1952~W1 and C/1978~G2,
see Table~4 in Paper~1 and Table~7 in Paper~2, respectively) similar
to C/2010~H1 with formally negative value of $1/a_{{\rm orig}}$
for the nominal osculating orbit but highly dispersed swarm with high
per cent of elliptical orbits. These four comets form the majority
of the negative tail of the global $1/a_{{\rm orig}}$-distribution
given in Fig.~\ref{fig:stat108_new_old_uncertain}. Some of the other
comets provide quite insignificant contribution to this negative tail.

\begin{center}
\begin{table}
\caption{\label{tab:previous_motion_dynamical_status}The percentage contributions
of dynamically new/old/uncertain parts of VCs to individual bins of
$1/a_{{\rm orig}}$-distribution for all 108~comets studied by us.
The range of $1/a_{{\rm orig}}$ corresponds to the area of histograms
given on a grey background in both panels of Fig.~\ref{fig:stat108_new_old_uncertain}.
Numbers in the upper part of the table are compatible with histograms
in Fig.~\ref{fig:stat108_new_old_uncertain}. In the lower part of
this table, we show only rows where the numbers were changed as a
result of another definition of uncertain VCs. }

\begin{tabular}{cccccc}
\hline 
Range of $1/a_{{\rm ori}}$  & \multicolumn{4}{c}{Per cent (in a bin) of dynamically} & Per cent \tabularnewline
\hline 
bin in units of  &  &  &  &  & of all VCs \tabularnewline
$10^{-6}$\,au$^{-1}$  & new  & old  & unc.  & old$+$unc.  & in a bin \tabularnewline
\hline 
 &  &  &  &  & \tabularnewline
\multicolumn{6}{c}{For the range of 'uncertain' VCs: 10\,au$\le q_{{\rm prev}}\le$20\,au}\tabularnewline
0--10  & 100  & 0  & 0  & 0  & 4.7 \tabularnewline
10--20  & 100  & 0  & 0  & 0  & 14.1 \tabularnewline
20--30  & 88.5  & 2.9  & 8.6  & 11.5  & 14.7 \tabularnewline
30--40  & 39.2  & 25.7  & 35.1  & 60.8  & 15.5 \tabularnewline
40--50  & 3.4  & 75.5  & 21.1  & 96.6  & 17.5 \tabularnewline
50--60  & 0  & 94.8  & 5.2  & 100  & 12.1 \tabularnewline
60--70  & 0  & 96.6  & 3.4  & 100  & 6.9 \tabularnewline
70--80  & 0  & 100  & 0  & 100  & 3.4 \tabularnewline
80--90  & 0  & 100  & 0  & 100  & 1.3 \tabularnewline
90--100  & 0  & 100  & 0  & 100  & 1.6 \tabularnewline
\hline 
 &  &  &  &  & \tabularnewline
\multicolumn{6}{c}{For narrower range of 'uncertain' VCs: 10\,au$\le q_{{\rm prev}}\le$15\,au}\tabularnewline
20--30  & 93.3  & 2.9  & 3.8  & 6.7  & 14.7 \tabularnewline
30--40  & 57.7  & 25.7  & 16.6  & 42.3  & 15.5 \tabularnewline
40--50  & 7.0  & 75.5  & 17.5  & 93.0  & 17.5 \tabularnewline
\hline 
\end{tabular}
\end{table}

\par\end{center}

\section{Summary and conclusions}

\label{sec:Summary}

This paper is a continuation of our pending effort to study past and
future motion of all observed LPCs for which a reliable osculating
orbit can be determined. During a few last years, we focus our attention
on so called Oort spike comets, i.e. actual comets with the largest
semimajor axes in the whole population of observed LPCs. Thus, the
investigated here dynamical motion of the sample of 22~near-parabolic
comets with small perihelion distance and discovered in 2006-2010
is a continuation of our long-standing studies. A detailed description
and statistics concerning osculating, original and future orbits of
these comets may be found in the first part of this paper \citep{kroli-dyb:2013}.
There is also a separate paper by Królikowska \citeyearpar{krolikowska:2014}
describing in detail a publicly available catalogue of near-parabolic
comets, which offer osculating, original and future orbits, augmented
with their uncertainties and the overall quality assessments already
introduced in Part~I.

\vspace{0.2cm}

Our main conclusions raised from this investigation of 22~comets
recently discovered are following: 
\begin{itemize}
\item Small perihelion comets discovered in 2006-2010 appear rather typical
when comparing with many other LPCs studied by us earlier. The only
exception is C/2007~W1 which seems to be the first  serious candidate for interstellar provenience.
\item Studying the past motion of these 22 comets, we found nine of them
to be dynamically new, eight dynamically old and five for which we
cannot determine their dynamical status according to the adopted criteria.
Among dynamically old comets, we found only one (C/2007~Q3) to have
a previous perihelion distance close to 10\,au, the rest have this
parameter smaller than 5\,au. One comet in the group investigated
here, C/2007~Q1 (very short data-arc span of 24\,days) has such
an uncertain osculating orbit that any conclusions about its provenience
seems impossible. 
\item The future motion of these comets exhibits various dynamical behaviour.
C/2010~X1 disintegrated during the observed perihelion passage, maybe
two other also disintegrated after perihelion passage, or split, however,
there is nothing for sure (for more details see Part~I). Six comets
will escape from the Solar system along hyperbolic orbits and fifteen
will return to the solar neighbourhood along much more tighten orbits
than their original ones. Again, for C/2007~Q1 we cannot make any
definitive conclusion due to a large uncertainty of its future orbit. 
\end{itemize}
If we compare this sample of 22 comets with 86 LPCs studied by us
earlier we can observe that: 
\begin{itemize}
\item The percentage of dynamically old comets in the investigated here
sample of small-perihelion comets is comparable to the remaining sample
of 86~comets analysed in Paper~I-II, where we found 38~comets (44
per cent) which are dynamically old and several with uncertain status.
The same is true for dynamically new comets. For the entire sample
of 108~comets we have 42 per cent of dynamically old comets and the
same per cent of dynamically new comets, about 16 per cent have uncertain
status. 
\item It seems that we have here no more than 30~per cent of comets which
leave the Solar system on hyperbolic orbits. The ambiguity associated
with the future of a few comets mentioned above, does not change this
picture. Thus, this sample seems to be quite different than the group
of small-perihelion comets analyses in Paper~I where we found the
opposite tendency -- there 60~per cent of comets will be lost on
hyperbolic orbits. However, for the sample of 108~comets we have
54~per cent of escaping comets on hyperbolic future orbits. But this
is effect of planetary perturbations (with some additional influence
of nongravitational effects) as it was described in Part I and here
we can only add that neither Galactic nor stellar perturbations change
the fate of any of LPCs studied by us. 
\item The dependence of perihelion distance change during the last orbital
period on the original semimajor axis seems to be similar in all samples
we studied, as it is shown in Fig.\ref{fig:Overall-dependence-13}. 
\item It is worth to mention here, that for large percentage of small perihelion
distance LPCs (50 per cent in this paper, all comets from Paper I)
and significant number of large perihelion distance comets (over 20
per cent from Paper II) we used nongravitational orbits. As we already
shown original and future semimajor axes of LPCs are often highly
influenced by nongravitational forces. This makes our results for
previous and next orbits much more reliable for this objects. 
\end{itemize}
Since the total number of near-parabolic comets studied by us so far
(108 objects) is quite large we can formulate some more general conclusions: 
\begin{itemize}
\item This paper again confirms our previous results that not all comets
from the Oort spike are dynamically new (see Fig.\ref{fig:stat108_new_old_uncertain}).
We observed that the significant percentage of near-parabolic in this
investigation as well as in our earlier studies have their previous
perihelia deep in the planetary region. As a result, one cannot treat
them as \emph{new comets} since they experienced both planetary perturbations
and solar radiation heating at least during their previous perihelion
passage. 
\item The second key conclusion is that the widely used concept of the Jupiter-Saturn
barrier should be revised since significant number of near-parabolic
comets (about 15 per cent) can migrate through it without any significant
orbital changes. As a result obtaining small previous perihelion distance
do not necessarily make a comet dynamically new, while its thermal
interaction with the Sun is obvious. This might be an important factor
since one can imagine several consecutive perihelion passages of near-parabolic
comets close to the Sun what complicates our understanding of dynamical
and physical ageing of LPCs. 
\end{itemize}
For the first time in our calculations we fully account  for perturbations from all known stars
therefore the second aim of this paper was to test their
significance. We have extended our test for all 108~comets investigated
by us so far. From the point of view of discriminating between dynamically
old and new comets we did not found any significant changes. Only
three comets (C/1996~E1, C/2001~K3 and C/2006~OF$_{2}$ are reclassified
into the uncertain category due to small previous perihelion changes.
As it is shown with examples in Section~\ref{sec:Examples-of-the}
the perturbations from all known stars are weak and often compensate
each other. There is only one star, Gliese~710 (HIP 89825) which
can make arbitrarily close encounter with LPCs in the future. We found
the extreme case for comet C/1999 F1 which nominally will approach
Gliese~710 as close as 11\,000\,au.

Our computer codes are fully prepared to use more potential stellar
perturbers so if the ongoing GAIA mission reveal stars or stellar
systems we missed so far, we can easily check their importance for
any particular near-parabolic comet.

 \bibliographystyle{mn2e}
\bibliography{moja23}

\label{lastpage} 
\end{document}